\documentstyle[referee]{l-aa}

\def\mod{\mathop{\rm mod}\nolimits}
\def\tvi(#1,#2){\vrule height #1pt depth #2pt width 0pt}
\def\hMpc{h^{-1}\,\mbox{Mpc}}
\begin{document}

%\thesaurus{?????}
\title{Cosmic crystallography}
\author{R.~Lehoucq\inst{1}, M.~Lachi\`eze--Rey\inst{1,2} \and
J.P.~Luminet\inst{3}}
\offprints{J.P.~Luminet}
\institute{CE-Saclay, DSM/DAPNIA/Service d'Astrophysique, F-91191 Gif sur
Yvette
cedex, France
\and
CE-Saclay, DSM/DAPNIA/Service d'Astrophysique, CNRS--URA 2052, F-91191 Gif sur
Yvette cedex, France
\and
D\'epartement d'Astrophysique Relativiste et de Cosmologie, CNRS--UPR 176,
Observatoire de
Paris--Meudon, France}
\date{september 1995}
\maketitle

\begin{abstract}
We assume that the Universe has a non trivial topology whose compact spatial
sections have a
volume significantly smaller than the horizon volume. By a topological lens
effect, such a
``folded'' space configuration generates multiple images of cosmic sources,
e.g.
clusters of
galaxies. We present a simple and powerful method to unveil non ambiguous
observational effects,
independently of the sign of the curvature and of the topological type. By
analogy with
techniques used in crystallography, we look for spikes in the pair separation
histogram between
cosmic objects in 3-D space. The spikes due to multi--connectedness should
stand
out
dramatically. Moreover, their positions and their relative amplitudes would be
definite
signatures of the topological type and of the underlying geometry. Such a
statistical method
would thus reveal the shape of space. As illustrative examples, we perform
numerical simulations
in $\Omega = 1$ Friedmann universes with the six possible closed orientable
topologies, which
prove the efficiency of our method. Presently available 3D catalogs of galaxy
clusters are not
deep enough to test our method at sizes greater than lower limits $\approx
600\,\hMpc$ previously
obtained by other methods. With extensive redshift surveys currently in
progress
the situation
may change in the next decade.


\end{abstract}

\section{Introduction}

It is presently believed that our Universe is correctly described by one of the
spatially
homogeneous and isotropic Friedmann--Lema{\^\i}tre models. Their spatial
sections are of the
elliptical, Euclidean or hyperbolic type according to the sign of their
constant
spatial
curvature. Most studies in the field assume that the topology of space is
simply--connected, but
the possibility of a non trivial topology becomes an increasingly popular topic
in theoretical
and observational cosmo\-logy (for a review, see Lachi\`eze--Rey and Luminet
1995, hereafter
LaLu95 and references therein; also Fagundes 1995).

For the simply--connected models, the finite or non finite character of space
is
linked to the
sign of spatial curvature: elliptical models have finite volumes whereas
Euclidean and
hyperbolic models have infinite volumes. In that case the question of the
extension of space
(one of the oldest cosmo\-lo\-gi\-cal problems, going back to almost twenty
five
centuries) is
reduced to the estimation of the average mass-energy density of the Universe
(including a
possible cosmo\-lo\-gi\-cal constant's contribution). The situation is
different
for
multi--connected models (Ellis 1971), since the three geometries of constant
curvature admit
compact space forms (although the elliptical family do not admit non compact
ones).

Present observations do not clearly indicate the value of the spatial curvature
of the
Universe, neither whether space is simply or multi--connected. However
observational effects are
expected if the physical space is compact in at least one direction and if its
corresponding
size is smaller than the horizon distance. We call such a space ``small
folded''. Two of us
(LaLu95) published a critical review of the various methods already proposed to
check the cosmic
topology; most of them present at least one of the fol\-lo\-wing drawbacks: (i)
they apply to
one particular type of multi--connectedness only, for instance the torus like
topology; (ii)
they rest on strong assumptions about the cosmological model, for instance an
Einstein--de
Sitter universe, or on the properties of a peculiar population of cosmic
objects, for instance
quasars; (iii) they do not provide an unambiguous signature of
multiconnectedness.

In this article we propose to test if space is a small folded one by studying
the histogram of
pair separations between cosmic sources. Our method is an improved version of
the construction
of the correlation function of clusters already pioneered by Schvartsman and
his
group
(references in LeLa95). It is free from the preceding drawbacks and is
independent of
preliminary assumptions about the topological type.

We recall general properties of multi--connected universe models in section~2.
We present the
pair separation histogram method in section~3. We check its validity on
simulated universe
models in section 4. We perform the test on a 3D catalog of galaxy clusters and
draw conclusions
in section 5.

\section{Small folded universes}

\subsection{General properties}

Mathematically, a space is simply--connected if every loop can be continuously
shrunk to a
point. If not, the space is multi--connected. A multi--connected space is
conveniently described
by its fundamental polyhedron $F$ and its holonomy group $\Gamma$. $F$ is
convex, with a finite
number of faces which are identified by pairs. $\Gamma$ is generated by
transformations which 
carry a face to its homologous one. The latter are isometries without fixed
point. The
fundamental polyhedron is transformed into its images $\gamma F$ by the
holonomies $\gamma \in
\Gamma$. The reunion of the $\gamma F$'s form a regular tiling of the
so--called
universal
covering space. A simply--connected space is identical to its universal
covering
space, whereas
a multi--connected space is the quotient of the universal covering space by the
holonomy group
of its fundamental polyhedron.

Applied to the spatial sections of cosmological models, the fundamental
polyhedron indeed
identifies with the physical space, in which objects like galaxies or clusters
are located,
whereas the universal covering space identifies to $\bbbs^3$, $\bbbr^3$ or
$\bbbh^3$ according
to the sign of the curvature (we assume, as usual in cosmology, homogeneity and
local isotropy
of space).

A given cosmic object lies at a given position in physical space, i.e. in $F$.
Its images under
$\Gamma$ lie in the universal covering space. The latter is thus the
``observer's world'', which
may drastically differ from the real world. As an example, the observer may see
several images
of the same cosmic object if $F$ is smaller than the particle horizon distance
in at least one space
direction. This gives rise to the appellation of ``small folded'' universe. By
convention we call the nearest source the ``original'' and the other images
``ghosts''.

Small folded universe models are easier to understand with the follo\-wing
remarks: (i)
although it is not the physical space, the universal covering space in a given
multi--connected
Friedmann--Lema{\^\i}tre model has exactly the same properties than the
physical
space of the
corresponding simply--connected model; (ii) a ghost object in the universal
covering space of a
multi--connected model has the same properties (observed distance, redshift,
age) as the object
located at the same position in the real space of the corresponding
simply--connected model.

A multi--connected universe model is characterized by some spatial scales
associated to the
fundamental polyhedron. Let us call $\alpha$ its smallest length. In a non flat
space,
characterized by its present curvature radius $R_0$, the ratio $\alpha/R_0$ is
geometrically
constrained to specified values (LaLu95). For instance it has a maximum (resp.
minimum) value if
$k > 0$ (resp. $k < 0$), whereas it remains arbitrary in flat space. The
fundamental polyhedron
also involves another scale $\beta$, the maximum length inscriptible in it
(this
is for
instance the diagonal for a parallepipedic fundamental polyhedron, which
characterizes the
hypertorus). This is also the maximum distance between 2 images of the same
object belonging
to adjacent cells.

Directly observable effects are expected if $\alpha$ or $\beta$ are smaller
than
the horizon
size. By construction, only original images (no ghosts) are present up to a
distance $\alpha
/2$. Between $\alpha /2$ and $\beta /2$ there is a mixture of original and
ghost
images. Beyond
$\beta /2$ there are only ghosts. Of course, the smaller the basic cell, the
easier to detect
topological effects. The number of potential ghosts depends on the cosmic
parameters and on the
topology. For a given object, there are as much potential ghosts as cells in
the
universal
covering space; their number is thus finite in the case of positive constant
spatial curvature,
infinite otherwise. An obvious limitation is that only ghosts nearer than the
particle horizon
may be seen, so that their number is at most equal to the number of cells
within
the particle's
horizon. In addition, the number may be drastically reduced by the fact that in
practice, we can
only observe below some magnitude or redshift cut-off, depending on the type of
objects
(galaxies, clusters, etc.), on the instrumentation and various other
limitations.

\subsection{Observational constraints}

No direct observation presently indicates that our universe is a small folded
one. By direct
observation we mean recognition of ghost images of known cosmic objects or
configurations. This
places lower limits on the size of the physical space according to the type of
objects. Some tests
involve the diffuse cosmic microwave background, other the discrete sources
such
as quasars or
clusters. From the COBE/DMR results, Stevens et al. (1993), Starobinsky (1993)
and de
Oliveira-Costa \& Smoot (1995) have claimed to rule out multi-connectedness on
sub-horizon scales.
However their results rely onto various disputable assumptions, for instance
the
absence of strong
reionization after recombination (see section 12.3 in LaLu95 for an extended
discussion). Fagundes
(1995) tried to fit the cold and hot spots in COBE/DMR maps of the cosmic
microwave background
with the predictions of some compact hyperbolic models with $\Omega < 1$,
whereas Jing and Fang
(1994) fitted the two-point angular correlation function of the cosmic
microwave
background. At
present time, limits derived from the cosmic microwave background observations
are not model
independent and cannot be considered as definitive yet. Thus it remains
valuable
to check the
cosmic topology with a different method.

Tests which involve discrete sources must consider populations of objects
extending deep enough
in space to check large dimensions. Quasars seem potentially interesting, but
their estimated
lifetime is short compared to the expected time necessary for a light ray to
turn around a small
folded universe: a typical lifetime less than $\approx 10^8$~yrs would only
allow to
investigate scales smaller than 200~Mpc.

The observations of clusters and superclusters of galaxies seem more adapted.
They have led to the
limits $\alpha > 60\,\hMpc$ (Gott 1980) and $\beta > 600\,\hMpc$ (Sokoloff and
Schvartsman 1974, Fang
\& Liu 1988), where $h$ is the Hubble constant in units of
100~km~sec$^{-1}$~Mpc$^{-1}$. To our
knowledge, they are the best observational constraints derived from discrete
sources. They leave room
for many observable effects in a folded universe. To get a qualitative idea,
let
us assume an
Einstein--de Sitter model (zero curvature) with a cubic hypertorus of length
$L$
as fundamental cell.
The redshift--distance relation is $d(z) = d_H\,\left(1 - (1 +
z)^{-1/2}\right)$, where $d(z)$ is the
distance in the universal covering space, and $d_H = 2 c/H_0 \approx
6000\,\hMpc$ is the horizon
distance. Table~1 gives the multiplication factor, namely the number of ghost
images of a given
original object, for various cell sizes compatible with present observations.

\begin{table}
\caption[]{Multiplication of images in a cubic hypertorus universe with
fundamental lengths
$L$, in units of $\hMpc$. $z_{\alpha}$ is the redshift until which there are
only originals,
$z_{\beta}$ the redshift beyond which there are only ghosts, $M(d_H)$ the
multiplication
factor until the horizon distance $d_H$, $M(z)$ the multiplication factors
until
various
redshift cut-offs. Magnitude limitations, absorption, luminosity evolution and
other
observational biases are neglected.}
\begin{center}
\begin{tabular}{|c|c|c|c|c|c|c|}
\hline
$L$& $z_{\alpha}$& $z_{\beta}$& $M(d_H)$& $M(4)$& $M(1)$& $M(0.5)$ \\ \hline
 500& 0.09& 0.16& 7000& 1200& 180& 45  \\ \hline
1000& 0.19& 0.37&  900&  150&  23&  5  \\ \hline
1500& 0.31& 0.63&  279&   45&   7& 1.5 \\ \hline 
2000& 0.44& 0.98&  110&   20&   3&  -- \\ \hline 
2500& 0.60& 1.45&   60&   10& 1.5&  -- \\ \hline
\end{tabular}
\end{center}
\end{table}

The limits mentionned above are mainly derived from the absence of ghosts of
peculiar objects
(the Milky Way, some known clusters or superclusters). However some ghosts
would
be unobservable
for various reasons not linked to the geometry (absorption by dust, difficulty
to recognize the
object, etc.). This motivates a search for statistical tests which would be
less
dependent on
peculiar objects and would exploit the largest amount of the information
contained in 3-D
catalogs.

Different methods have been proposed, such as the search for a periodic or
quantized
distribution of objects in redshift or in distance (Fang 1990). For instance,
it
may be
tempting to invoke multi--connectedness for explaining the apparent periodicity
observed by
Broadhurst et al. (1990) in a pencil-beam catalog of galaxies. But Park and
Gott
(1991), among
others, were able to reproduce comparable results in ordinary simply--connected
universe models.
As emphasized in LaLu95, such tests cannot be decisive, and our numerical
simulations below
confirm this point of view.

\section{A pair separations histogram for testing cosmic topology}

Since the above methods do not provide a clear signature of
multi--connec\-ted\-ness, we propose
a more general one, independent on the peculiar topology or on the peculiar
cosmic objects. In this
purpose, we go back to the basic property of folded universes: each image of a
given object is
linked to each other one by the holonomies of space. Without knowing the
geometry or the topology
we do not know what are these holomies, but we know that they are isometries.
Therefore, any pair
of images comprising an original and one of its ghosts, or two ghosts of the
same object,
reflects an isometry of space. We call such a pair a gg-pair. A signature of
multi--connectedness
would be that, among all pairs of images, a significant proportion are
gg-pairs.
So the question
is: how to extract gg-pairs from ordinary pairs~? The answer is provided by the
histogram of
space separations between all pairs of images in a 3--D catalog, as we see now
in more detail.

A topology is characterized by its holonomies, which are combinations of the
generators
$\{\gamma_k\}$ of the holonomy group. Each generator $\gamma_k$ is itself an
holonomy, to which
is associated an identification length $\lambda_k$ related to the size of the
fundamental
polyhedron. For instance, in toroidal models the $\lambda_k$ are equal to the
edges $L_e$, $L_a$
and $L_u$ of the parallepipedic fundamental polyhedron. More generally, the
$\lambda_k^2$ are
related to $L_e^2$, $L_a^2$ and $L_u^2$ by linear relations involving integer
coefficients.

All the gg--pairs corresponding to the same generating holonomy $\gamma_k$ are
characterized by
the same separation $\lambda_k$ (in proper comoving distance units). Thus, if
we
draw the
histogram of the pair separations, or rather the squared separations, these
gg--pairs will emerge
from ordinary pairs as a spike located at the position $\lambda_k^2$. In
addition to the
generating holonomies, other holonomies appear as compositions of them, also
giving
characteristic spikes. They also generate gg--pairs with squared separations
\begin{equation}
\label{entire}
\Lambda _i^2 = \sum N_k \lambda_k^2,
\end{equation}
where the $N_k$ are integers. 

The principle of our test is to recognize the presence of spikes associated to
these values, in a
catalog of observed cosmic sources. Let us assume that our catalog of $N$
objects (thus containing
$N(N-1)/2$ pairs) has a characteristic volume about $F$ times that of the
fundamental polyhedron. The
latter thus contains about $N/F$ (original) objects, the other being ghosts.
Each ghost $g_1$
(excepted near the edges) is transformed by the holonomy generator $\gamma_k$
to
give an other ghost
$g_2$, their separation being $\lambda_k$. Thus, we expect about $N$ (in fact
less because of edge
effects) corresponding gg-pairs with separation $\lambda_k$, which will emerge
as a spike above the
background contribution of ordinary pairs. If 2 or 3 identification lengths are
equal, this number is
to be multiplied by 2 or 3 and the corresponding spike will be enhanced.

In a small folded universe, the spikes would be observable and provide a
signature of
multi--connectedness, like in a crystallographic lattice. The integer values
$N_k$ would be
measurable and (together with the relative heights of the spikes) indicate
exactly the type of
topology involved. This is illustrated in the next section. Note that the
practical calculus of
separations depending on the curvature of space, the analysis has in principle
to be performed
in the three different cases $k = -1,0,1$. However, the Euclidean approximation
applies well when
the catalog does not extend too deeply.

\section{Simulated folded universes}

\subsection{Locally Euclidean models}

In order to check the validity of our test, we have generated simulated
catalogs
of a
distribution of objects (typically, galaxy clusters or quasars) in a
multi--connected universe.
For simplicity, we chose a flat Friedmann--Lema{\^\i}tre (Einstein--de Sitter)
universe, whose
universal covering space is $\bbbr^3$. We investigated the six compact oriented
topologies,
denoted ${\cal T}_1$ to ${\cal T}_6$ (see Wolf 1984, and Fig.~17 of LaLu95 for
a
pictorial
representation). Each type is specified by a fundamental polyhedron with three
edge lengths
$L_e$, $L_a$, $L_u$, and the structure of its holonomy group. The latter may be
characterized by
the formulae identifying the coordinates ${\bf r} = (x, y, z)$ and ${\bf r'} =
(x', y',z')$ of a
gg--pair~:

\noindent for types ${\cal T}_1$, ${\cal T}_2$, ${\cal T}_3$
\begin{equation}
\label{t1t2t3}
{\bf r'} = R_{Oz}(\alpha_i)\,{\bf r}
+
\left(\begin{array}{c}
      n_e\,L_e\\n_a\,L_a\\ n_u\,L_u
      \end{array}
\right)
\end{equation}

\noindent for types ${\cal T}_4$
\begin{equation}
\label{t4}
{\bf r'} = R_{Ox}(n_e\pi)R_{Oy}(n_a\pi)R_{Oz}(n_u\pi)\,{\bf r}
+
\left(\begin{array}{c}
      n_e\,L_e\\n_a\,L_a\\ n_u\,L_u
      \end{array}
\right)
\end{equation}

\noindent for types ${\cal T}_5$, ${\cal T}_6$
\begin{equation}
\label{t5t6}
{\bf r'} = R_{Oz}(\alpha_i)\,{\bf r}
+
\left(\begin{array}{ccc} 
      1&      -1/2&0\\
      0&\sqrt{3}/2&0\\
      0&         0&1
      \end{array}
\right)
\left(\begin{array}{c}
      n_e\,L_e\\n_a\,L_a\\ n_u\,L_u
      \end{array}
\right)
\end{equation}

\noindent where $R_{Ox}(\theta)$, $R_{Oy}(\theta)$ and $R_{Oz}(\theta)$ are the
rotation matrices
of angle $\theta$ about the three coordinates axis, $n_e$, $n_a$ and $n_u$ are
integers denoting
the position of a cell in the universal covering space, $L_e$, $L_a$, $L_u$ and
$\alpha_i$ are
given in Table~2.

\begin{table}
\caption[]{The six locally Euclidean, closed, oriented 3-spaces}
\begin{center}
\begin{tabular}{|c|l|c|}
\hline
Type ${\cal T}_i$& Fundamental polyhedron& $\alpha _i$ \\  \hline
${\cal T}_1$& parallelipiped, $L_e$, $L_a$, $L_u$& $0$          \\ \hline
${\cal T}_2$& parallelipiped, $L_e$, $L_a$, $L_u$& $\pi\,n_u$   \\ \hline
${\cal T}_3$& parallelipiped, $L_e=L_a$, $L_u$   & $\pi/2\,n_u$ \\ \hline
${\cal T}_4$& parallelipiped, $L_e$, $L_a$, $L_u$&      --      \\  \hline
${\cal T}_5$& hexagonal prism, $L_e=L_a$, $L_u$   & $2\pi/3\,n_u [\mod 3]$ \\
\hline
${\cal T}_6$& hexagonal prism, $L_e=L_a$, $L_u$   & $2\pi/6\,n_u [\mod 6]$ \\
\hline
\end{tabular}
\end{center}
\end{table}

Type ${\cal T}_1$ corresponds to the well known hypertorus. Types ${\cal T}_3$,
${\cal T}_5$ and
${\cal T}_6$ constrain two lengths to be equal. The volume of a cell (also of
the physical
space) is $V = L_e\,L_a\,L_u$ for types ${\cal T}_1$--${\cal T}_4$, and $V =
3\sqrt{3}/2\,
L_e^2L_u$ for types ${\cal T}_5$ and ${\cal T}_6$.

\subsection{Analyses of simulated catalogs}

We randomly distribute 50 cosmic objects within the fundamental cell and we
calculate the ghost
images in the universal covering space, to simulate the appearance of the sky
up
to a redshift
cut-off $z = 4$. For galaxy clusters, we neglect the peculiar motions which
could slightly
deviate the positions of ghosts from a rigorous distribution.

In a first series of simulations we have chosen equal edges $L_e = L_a = L_u$,
adjusted so that
the fundamental cell has a volume $(1500\,\hMpc)^3$. According to Table~1, the
number of ghosts
is about 45 times the number of originals. Ghosts appear at $z > 0.31$ and
originals disappear at
$z > 0.63$. Figure~1 shows the 2--dimensional appearance of the sky. Fake
large-scale structures
are generated but, as expected, no information about multi--connectedness
appears.
\begin{figure*}
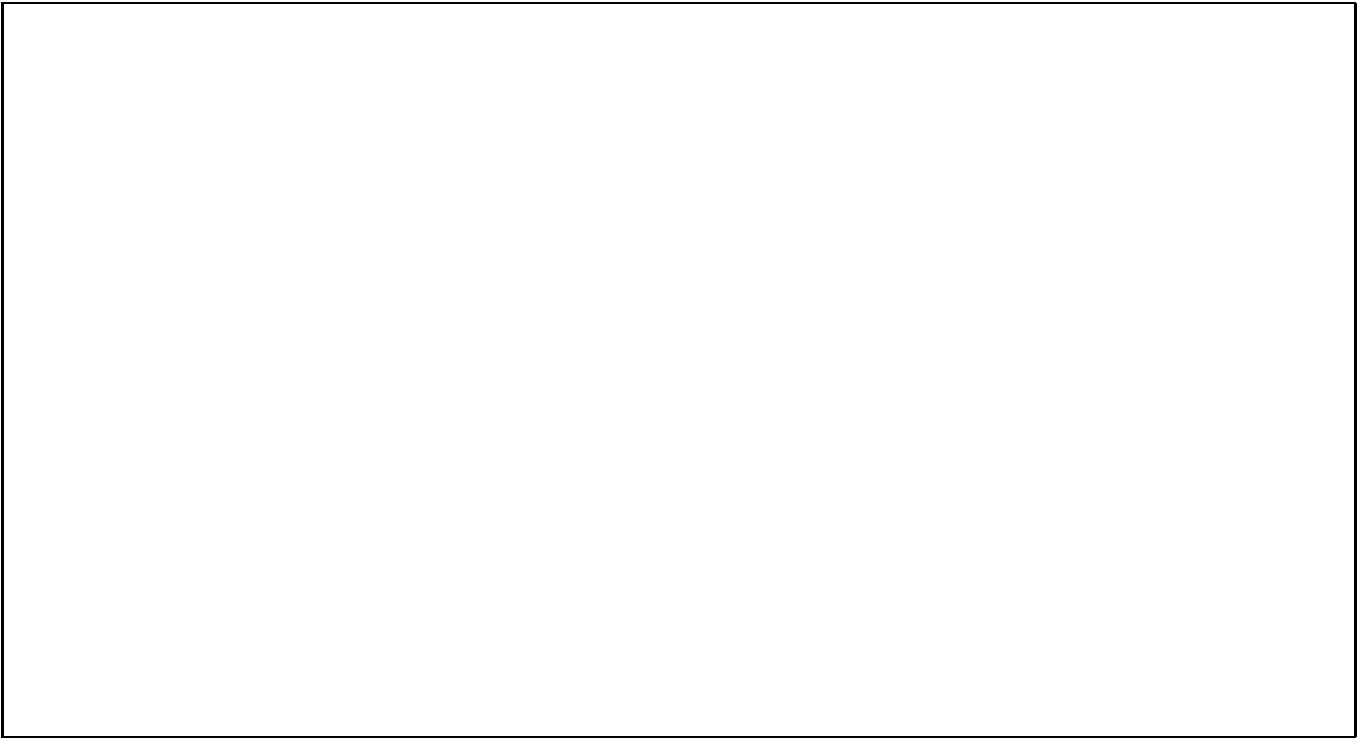

\picplace{9.75cm}
\caption[]{Appearance of the sky (equal area projection) in an Einstein--de
Sitter universe with
topology ${\cal T}_1$. The fundamental polyhedron is a cubic hypertorus whose
size is
$1500\,\hMpc$.}
\end{figure*}

We have also plotted in Fig.~2 the histogram of redshifts, calculated for the
hypertorus model.
No periodic pattern nor any sign of multi--connectedness appears. This confirms
former
simulations by Ellis and Schreiber (1986) and the statements by LaLu95
according
to which such
histograms cannot provide a valuable test for cosmic topology.
\begin{figure}
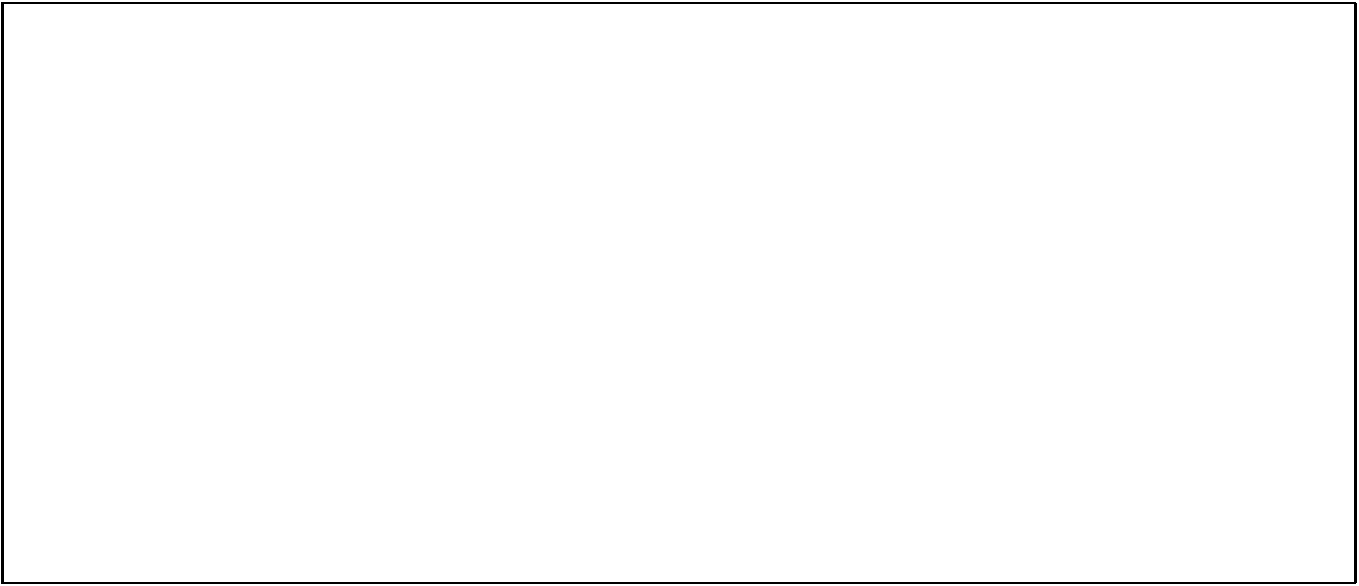

\picplace{7.7cm}
\caption[]{The redshift histogram generated by 50 randomly distributed objects
in the
fundamental cell of a ${\cal T}_1$ universe.}
\end{figure}

Next, for each simulated catalog of objects we have drawn the pair separation
histogram. They are
displayed in Fig.~3, from which the information about
multi--con\-nec\-ted\-ness
and the
topo\-lo\-gi\-cal type spec\-ta\-cu\-lar\-ly springs out.
\begin{figure*}
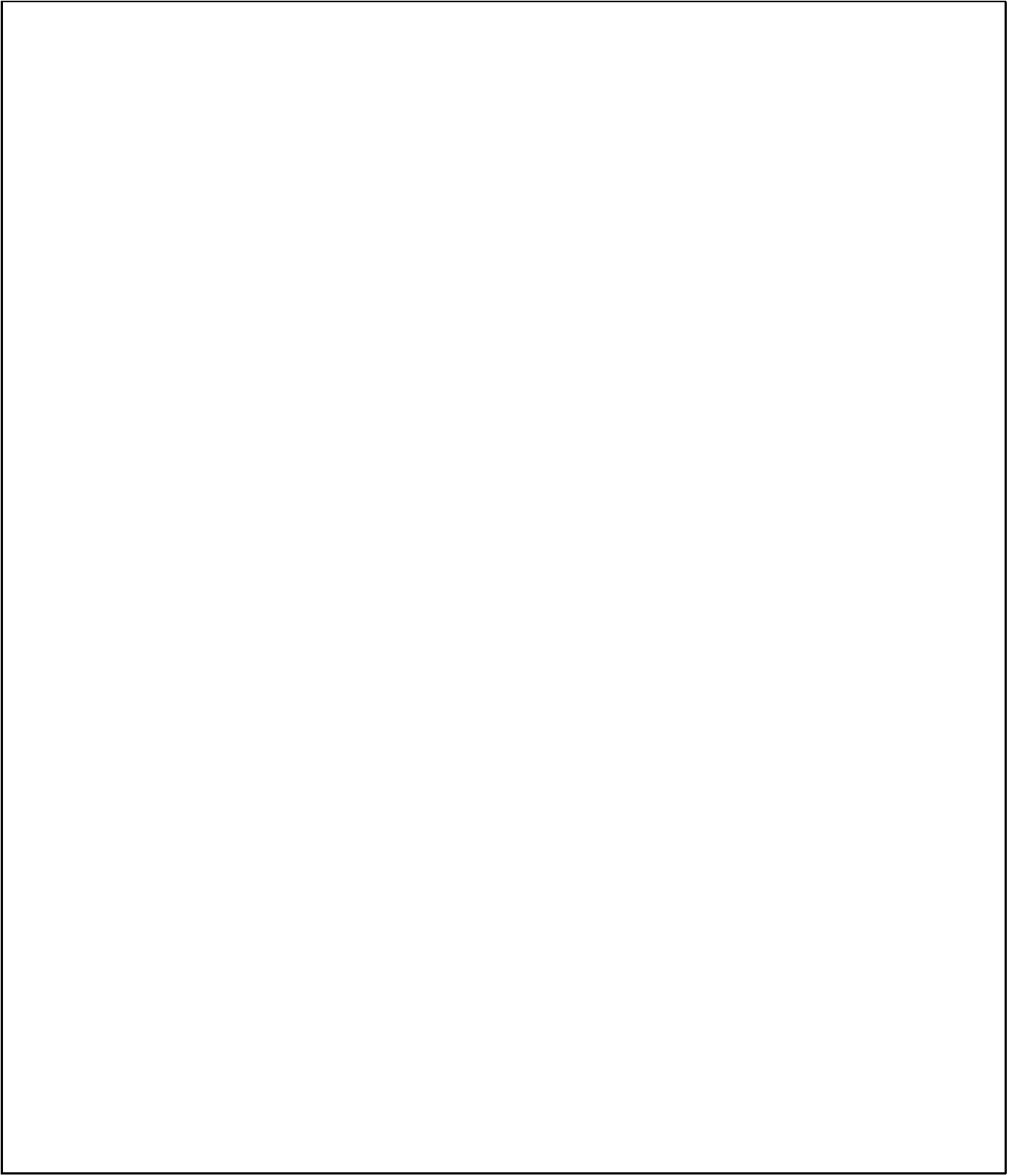

\picplace{21cm}
\caption[]{Histogram of squared separation distances between all pairs of
images
for the 6
topological types, with equal lengths. The spikes reveal repetition scales
related to the size
of the fundamental polyhedron.}
\end{figure*}

Spikes, which gather all gg--pairs linked by holonomies, ap\-pear at squared
sepa\-ra\-tions
$\Lambda_i$ obeying to Eq.~(\ref{entire}). A straightforward calculation of the
separations between
$(x',y',z')$ and $(x,y,z)$ according to Eqs.~(2--4) and Table~2 shows that
Eq.~(1) may be rewritten
under the very simple form
\begin{equation}
\label{nnn}
n_e^2 + n_a^2 + n_u^2 = \frac{\Lambda_i^2}{V^{2/3}}
\end{equation}
for the triplets on integers $(n_e,n_a,n_u)$ compatible with the topology. The
values of
$n$ span a range limited by the finite size of the catalog (0 to 5 in our
calculation). For a
given topology the positions of spikes are displayed in Table~3.

\begin{table}
\catcode`\*=\active \def*{\hphantom{0}}
\def\num{\hphantom{$^0$}}
\caption[]{The spike's spectrum of small folded Eucidean universes. The rank of
the three stronger
spikes are denoted by the exponant.}
\begin{flushleft}
\begin{tabular}{|c|c|c|c|c|c|}
\hline
\tvi(9,0)${\cal T}_1$&${\cal T}_2$&${\cal T}_3$&${\cal T}_4$&${\cal
T}_5$&${\cal
T}_6$ \\ \hline
\tvi(9,0)*1 \num&*1 \num&*1 $^2$&-- \num&-- \num&-- \num \\ \hline
\tvi(9,0)*2 $^3$&*2 \num&*2 $^3$&-- \num&-- \num&-- \num \\ \hline
\tvi(9,0)*3 \num&-- \num&-- \num&*3 $^1$&*3 $^1$&*3 $^1$ \\ \hline
\tvi(9,0)*4 \num&*4 $^3$&-- \num&*4 $^2$&-- \num&-- \num \\ \hline
\tvi(9,0)*5 $^1$&*5 $^1$&*5 $^1$&-- \num&-- \num&-- \num \\ \hline
\tvi(9,0)*6 $^2$&*6 \num&-- \num&-- \num&-- \num&-- \num \\ \hline
\tvi(9,0)-- \num&-- \num&-- \num&-- \num&*7 \num&*7 $^3$ \\ \hline
\tvi(9,0)*8 \num&*8 \num&*8 \num&*8 $^3$&-- \num&-- \num \\ \hline
\tvi(9,0)*9 \num&*9 $^2$&*9 \num&-- \num&*9 $^2$&*9 $^2$ \\ \hline
\tvi(9,0)10 \num&10 \num&10 \num&-- \num&-- \num&-- \num \\ \hline
\tvi(9,0)11 \num&-- \num&-- \num&11 \num&-- \num&-- \num \\ \hline
\tvi(9,0)12 \num&12 \num&-- \num&12 \num&12 $^3$&12 \num \\ \hline
\tvi(9,0)13 \num&13 \num&13 \num&-- \num&13 \num&13 \num \\ \hline
\tvi(9,0)14 \num&14 \num&-- \num&-- \num&-- \num&-- \num \\ \hline
\tvi(9,0)-- \num&-- \num&-- \num&-- \num&-- \num&-- \num \\ \hline
\tvi(9,0)-- \num&16 \num&-- \num&-- \num&16 \num&16 \num \\ \hline
\tvi(9,0)17 \num&17 \num&17 \num&-- \num&-- \num&-- \num \\ \hline
\tvi(9,0)-- \num&-- \num&-- \num&-- \num&18 \num&-- \num \\ \hline
\tvi(9,0)-- \num&-- \num&-- \num&-- \num&19 \num&19 \num \\ \hline
\tvi(9,0)-- \num&-- \num&-- \num&-- \num&-- \num&-- \num \\ \hline
\end{tabular}
\end{flushleft}
\end{table}

For a better understanding of Table~3, let us comment in more details the
spectrum of spikes
obtained for types ${\cal T}_1$ (cubic hypertorus) and ${\cal T}_4$. For ${\cal
T}_1$, the
holonomies are generated by the three translations along the edges of the
parallepipedic cell,
with magnitude $L = V^{1/3}$. In such a simple lattice structure, all the
triplets of integers
$(n_e, n_a, n_u) \in \{0,\dots,5\}$ realize an holonomy. Spikes thus appear at
all locations
$\Lambda_i$ such that $\Lambda_i^2/V^{2/3}$ is the sum of three squared
integers. Most integers
except 7, 15, 16\dots\ are solutions. For ${\cal T}_4$, the holonomies are
generated by two
translations and a glide reflection. For such a lattice structure, only a few
triplets of integers
$(n_e,n_a,n_u)$ realize an holonomy, so that Eq.~(\ref{nnn}) has much less
solutions.

Let us also examine the amplitude of the spikes. For type ${\cal T}_1$ for
instance, the spike at
position ${\Lambda_i^2}/V^{2/3} = 1$ is generated by the contributions of the
triplets $(1,0,0)$,
$(0,1,0)$ and $(0,0,1)$, whereas the spike at position $ {\Lambda _i^2}/V^{2/3}
= 5$ is generated
by the triplets $(2,1,0)$, $(2,0,1)$, $(1,2,0)$, $(1,0,2)$, $(0,1,2)$ and
$(0,2,1)$. Thus, spike 5
is more intense than spike 1. All the characteristics of Table~3 can be
described in a similar
way. We conclude that the very existence of spikes reveals the multi-connected
nature of topology,
whereas their positions and amplitudes discriminate the topological types.

We also performed the simulation in a toroidal model like ${\cal T}_1$, but
with
une\-qual
identification lengths. Figure~4 shows the result for $L_a$, $ L_e =
\sqrt{2}\,L_a$, and $L_u =
\sqrt{3}\,L_a$, so that $V = (1500\,\hMpc)^3$. The lengths being not
commensurable, the spikes are
more numerous, corresponding to all positions fundamental lengths and there
multiples, but less
intense since no gg--pair corresponding to different $\Lambda _k$ accumulate in
a same spike.
\begin{figure}
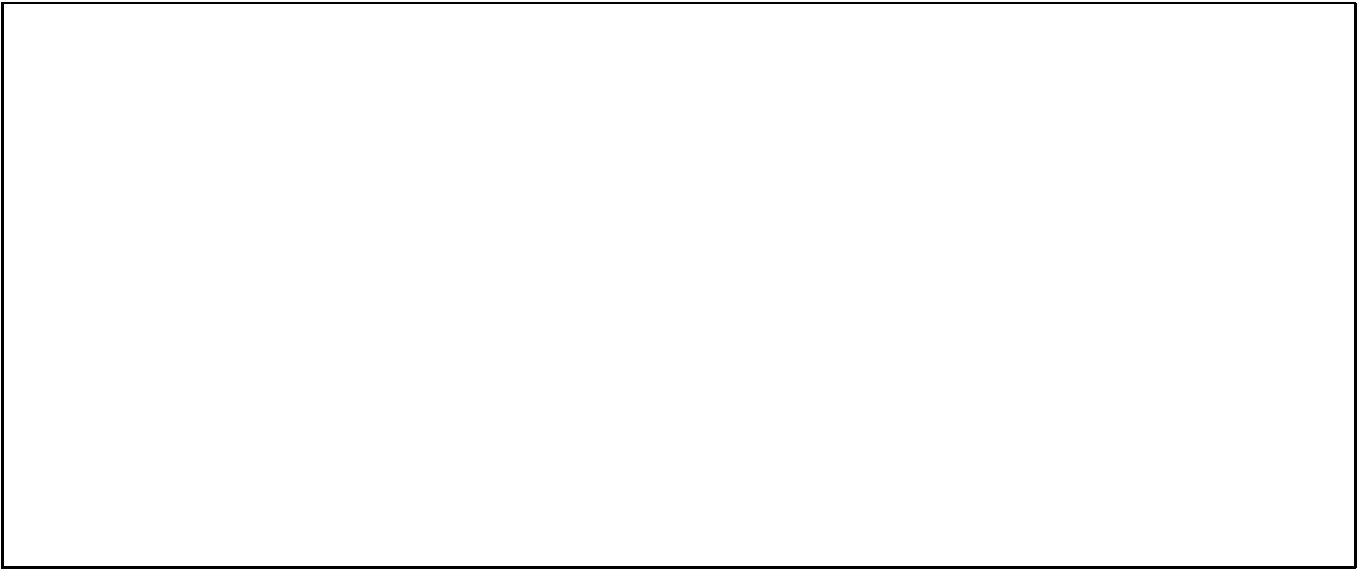

\picplace{7.5cm}
\caption[]{Histogram of squared separation distances for a ${\cal T}_1$
universe
with unequal
lengths.}
\end{figure}

Our simulated catalogs are however highly idealized so that the spikes due to
gg-pairs stand out
dramatically above the background distribution of other pairs. In a more
realistic distribution,
it would be more difficult to recognize the spikes, since various effects would
contribute to
spoil the sharpness of the spikes and decrease the signal to noise ratio. For
instance, the real
catalogs do not include the region masked by the galactic plane, and cover only
a small area of
the sky. In order to examine the consequences of this, we also created a
catalog, in a simulated
small folded universe, which is of limited solid angle. We observe that the
signal fades out when
the aperture angle of the catalog goes down to about $20\degr$ (see Fig.~5).
\begin{figure}
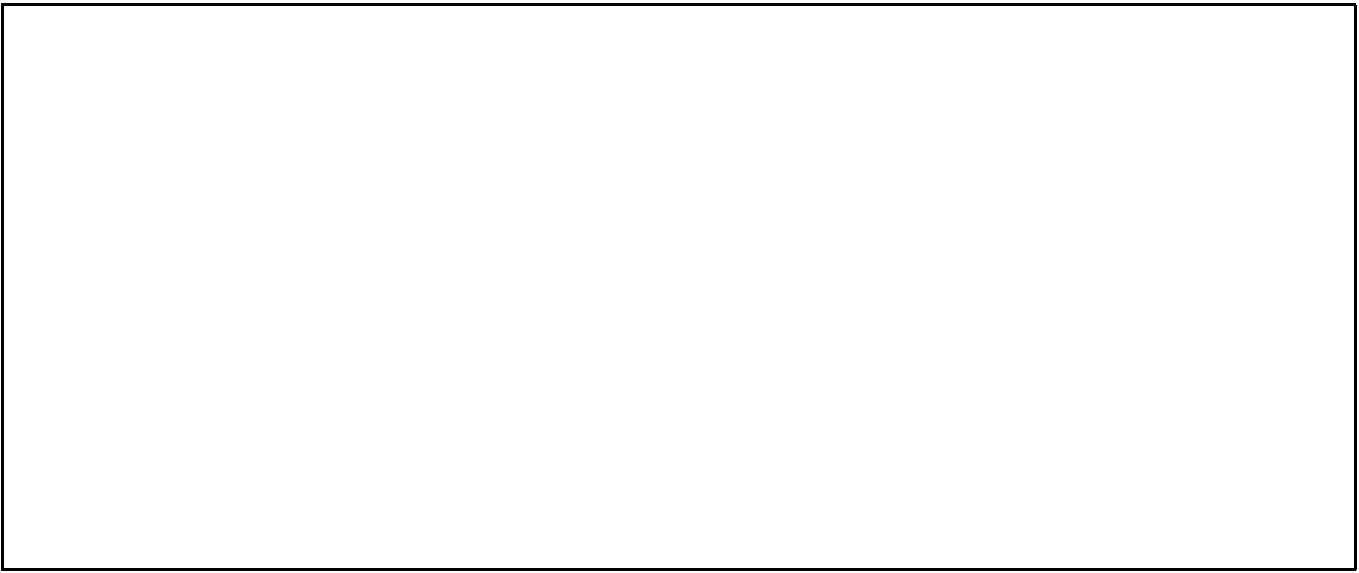

\picplace{7.5cm}
\caption[]{Histogram of squared separation distances inside a conical catalog
of
aperture
angle $20\degr$, for a ${\cal T}_1$ universe with equal lengths.}
\end{figure}

Also, clustering in a simply--connected universe can generate noise spikes that
mimic spikes due
to multi--connectedness. For instance, the spikes in the pair separation
histograms shown in the
Broadhurst et al. sample (1990) and in the Parks and Gott (1991) simulations
are
due to noise.
More precisely they are caused by the large number of pairs relating the many
galaxies in a
first cluster to those in a second cluster, all at the cluster separation.
Generally, such
spikes currently also occur in N-body simulations that show clustering.

\section{Results and discussion}

\subsection{Analysis of a cluster catalog}

We have applied our test to a 3--dimensional catalog of galaxy clusters
compiled
by Bury (private
communication). This catalog includes all Abell and ACO clusters with published
redshifts. It
contains 901 clusters, up to a maximal redshift $z_{max} \approx 0.35$,
corresponding to
$840\,\hMpc$ in an Einstein--de Sitter universe (although only 12 objects have
$z > 0.26$). Due
to obscuration by the galactic plane, the shape is a double cone of aperture
$\approx 120\degr$,
sufficient to allow the desired effects to be observable. Figure~6 shows its
projection onto the
sky.
\begin{figure*}
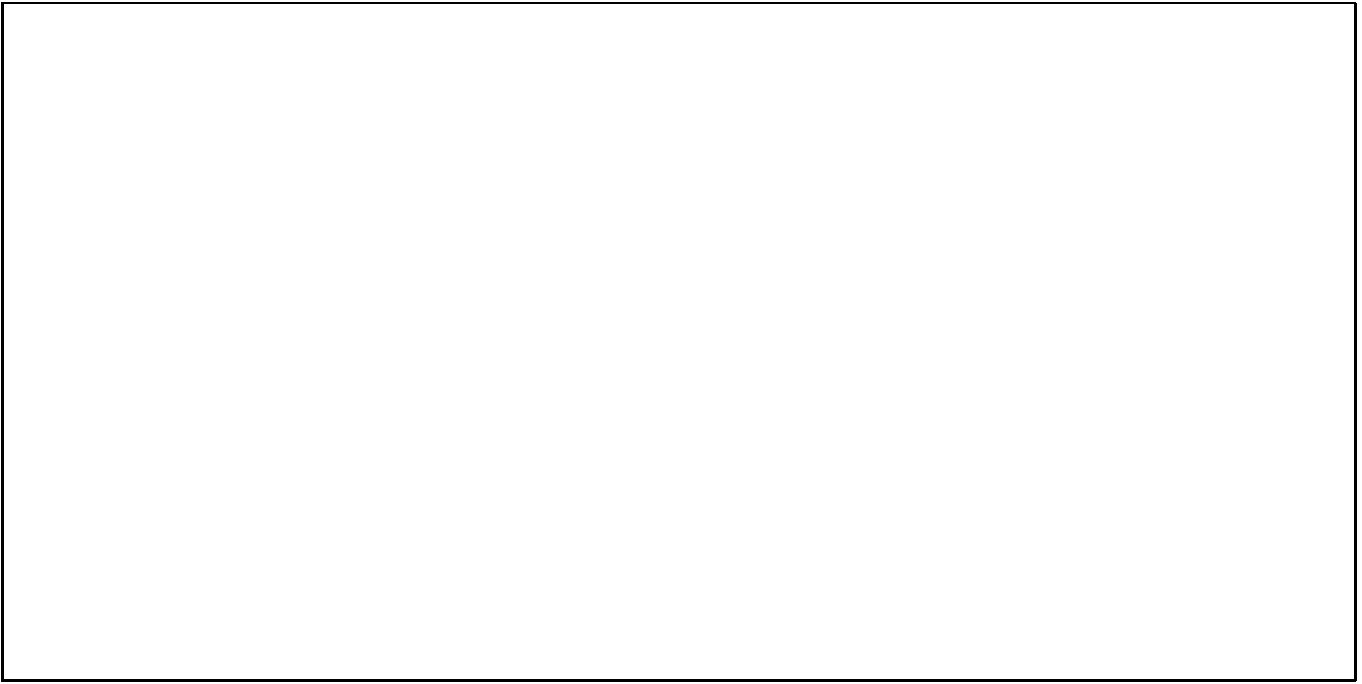

\picplace{9cm}
\caption[]{Two dimensional equal area projection onto the sky of the Bury
cluster catalog.}
\end{figure*}

\begin{figure}
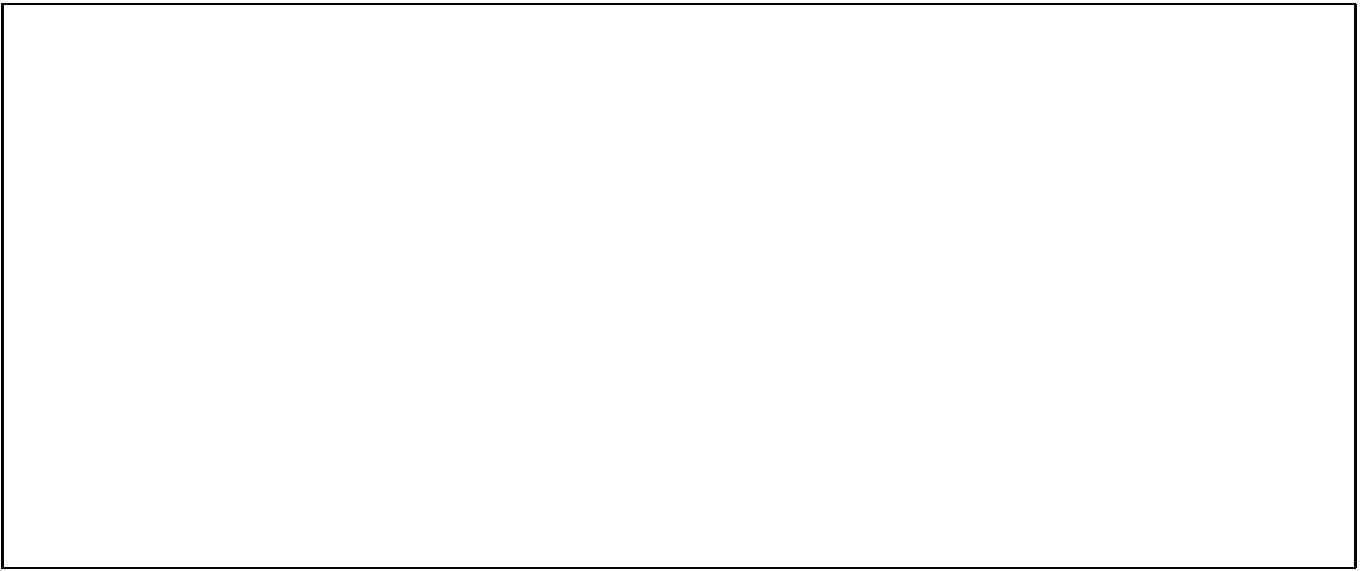

\picplace{7.5cm}
\caption[]{Histogram of squared separation distances for the Bury cluster
catalog.}
\end{figure}
Figure~7 presents the histogram of separations where two suspicious spikes are
detected corresponding
to separation of $270\,\hMpc$ and $382\,\hMpc$. The ratio of these quantities
is
very near
$\sqrt{2}$. If those spikes are really due to topology, this means that we are
seeing multiple images
of some given objects. In order to check this hypothesis, we have first
simulated the Bury catalog:
we distributed ramdomly 30 objects in a fundamental cell with $L_e = L_a = L_u
=
270\,\hMpc$, put a
redshift cut-off at $z = 0.26$, and limited our set of ghosts by a double cone
of aperture $120\degr$.
The number of real objects is chosen to obtain 901 ghosts in the final catalog.
Then, we plot the
histogram of separation on Fig.~8.

The comparison with Figs.~7 leaves no room to doubt. Multi--connectedness at a
scale $\approx
270\,\hMpc$ would give a much stronger signal than observed. To check further,
we also selected all
pairs contributing to the observed spikes. Were they due to
multi--connectedness, then the associated
separation vectors would clump in preferred directions, corresponding to the
principal directions.
Clearly, this is not the case as indicated by Fig.~9. Thus the observed spikes
are only due to noise
and no effect of multi--connectedness appears. Given the depth of the catalog,
we conclude to the limit $\alpha > 650\,\hMpc$, comparable to those already
obtained from large scale objects distribution. However applications of our
test
will be decisive when deeper 3D catalogs are available. Current observational
programs devoted to extended redshift surveys will offer this possibility
within
the next decades.

\subsection{Conclusion}

We have presented a new method to check and characterize a possible
multi--connectedness of space
on sub--horizon scales. Our test is free of specific assumptions about the
geometry and the
topological type. It is based onto the simple fact that a displacement carrying
a ghost image to
another of the same object is an isometry in the universal covering space.
Thus,
in a
multi--connected space, we expect that the histogram of all pair separations
between objects of a
large 3--dimensional catalog exhibits strong spikes, corresponding to
combinations of the
identification scales. Numerical simulations performed on Euclidean ``small
folded'' universe have
shown that the relative positions and amplitudes of the spikes characterize in
an unique way the
topological type: different topologies generate different recognizable spikes
structures.
Applying this test to present 3--dimensional catalog of galaxy clusters does
not
provide new
limits on the lower size of the universe.

Taking into account various effects such as luminosity evolution, peculiar
velocities and clustering of real objects within the fundamental cell would not
change the value of our test. In the future, it will be interesting to apply
the
test to more extended catalogs, and to collections of objects of different
types. Note also that the test can be made still more efficient by considering
pairs of specific configurations rather than isolated objects, like strings of
images with a given shape.

In the case of a significative distorsion, the pair separation histogram would
also provide a
powerful and purely geometrical method to determine the sign of the curvature
of
space. This
results from the fact that the crystallographic structures of the holonomy
groups differ for
the cases $k = 1, 0, -1$. Thus, beside its own specific interest,
multi--connectedness could
bring an answer to a long-standing problem of observational cosmology.

\acknowledgements{We thank the anonymous referee for useful comments on the
first manuscript and Bury for providing us his clusters catalog.}

\end{document}